# CMS (LHC) Measurements and Unusual Cosmic Ray Events

E. Norbeck and Y. Onel (for the CMS collaboration)
*University of Iowa, Iowa City, IA 52242, USA*

At the LHC, for the first time, laboratory energies are sufficiently large to reproduce the kind of reactions that occur when energetic cosmic rays strike the top of the atmosphere. The reaction products of interest for cosmic ray studies are produced at small angles, even with colliding beams. Most of the emphasis at the LHC is on rare processes that are studied with detectors at large angles. It is precision measurements at large angles that are expected to lead to discoveries of Higgs bosons and super symmetric particles. CMS currently has two small angle detectors, CASTOR and a Zero Degree Calorimeter (ZDC). CASTOR, at 0.7° down to 0.08°, is designed to study "Centauro" and "long penetrating" events, observed in VHE cosmic-ray data. As a general purpose detector it also makes measurements of reaction products at forward angles from p-p collisions, which provide input for cosmic ray shower codes. The ZDC is small, 9 cm. wide, between the incoming and outgoing beam pipes out at a distance of 140 m. The ZDC measures neutral objects that follow the direction of the beam at the interaction point. If the long penetrating objects are spectators they could be seen in the ZDC if their charge to mass ratio, Z/A, is less than 0.2.

## 1. INTRODUCTION

High altitude cosmic ray experiments find numerous examples of unusual events, whose nature is still not understood [1]. Of particular interest here are the "long flying" objects that have a penetrating power greater than ordinary hadrons. One category of such events goes under the name Centauro. The flux of Centauro events at the high altitude experiments at Chacaltaya and Pamir are substantial, of the order of 0.015 m$^{-2}$ yr$^{-1}$ [2]. With the LHC such events can be produced and studied in controlled accelerator experiments. The LHC is a colliding beam machine designed for p-p at 14 TeV and Pb-Pb at 1044 TeV (5.5 TeV/nucleon). The original design of the three major LHC experiments, CMS, ATLAS, and ALICE, features detailed measurements at large angles with respect to the beam direction. To observe the Centauro and related types of events requires detailed measurements at small angles. Because the nature of these objects is unknown, simple measurements are not sufficient for distinguishing the interesting events from more routine processes that will also be occurring. It would be best to use a tracking detector that could completely characterize the interaction of the unknown particle with the detector, but this would be difficult and expensive. The number of secondary particles in the resulting shower is large and the range of a Centauro is large; they are not significantly attenuated by 60 cm of Pb metal.

The CASTOR detector [3] was designed to demonstrate the existence of Centauro and related types of highly penetrating particles and measure their production cross sections without providing a detailed rendering of their interaction with the detector. CASTOR (Centauro And STrange Object Research) was constructed and added to the huge CMS experiment and has been collecting data since November, 2009. At present there is a CASTOR detector on only one side of CMS. It can be expected that the data gathered by the single CASTOR will be sufficiently provocative to justify a much larger expenditure for a second CASTOR with more detailed tracking capability.

## 2. CASTOR

CASTOR is a small angle detector covering the angles from 0.7° down to 0.08° (5.2<η<6.6) from the beam direction. Figure 1 shows the construction of CASTOR. Its total length provides 10.3 nuclear interaction lengths. It is divided into two parts, a short electromagnetic part (EM) with tungsten plate absorbers of 5 mm thickness and a long part (HAD) with also tungsten plates of 1.0 cm thickness for studying objects with the penetrating potential of hadrons. The showers are sampled by observing the Čerenkov light from thin quartz plates between the tungsten plates. The details of the longitudinal distributions are measured by photomultiplier tubes, PMTs that collect light from the quartz plates. There are 2 such reading units in the EM section and 12 in the HAD part. This pattern is repeated 16 times in the azimuthal direction for a total of (2 + 12) x 16 = 224 PMTs. It does not have any resolution in the radial direction. CASTOR is about 1.5 m in length and 36 cm in diameter and is about 16 m from the interaction point. It is divided into two halves, left and right, to allow installation against the beam pipe.

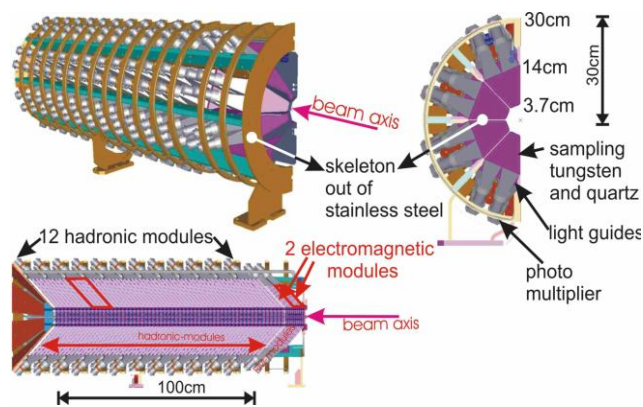

Figure 1: Details of the structure of CASTOR.

## 3. CENTAURO EVENTS

There are a wide variety of cosmic ray events with unusual characteristics. Of particular interest are the Centauro and related types. These are characterized by a large baryon to meson ratio with a range larger than hadrons and with a stopping power curve showing a dozen or so maxima while passing through 60 cm of Pb. It is thought that they are products of heavy nuclei, such as iron, reacting with atmospheric nitrogen or oxygen. Whatever they are, they most likely should be produced at LHC energies by Pb-Pb reactions. They will be identified by their characteristic stopping curves generated in one of the 16 azimuthal segments of CASTOR and by the EM to HAD ratio. Simulations [4] have shown that these events could easily be distinguished from statistical fluctuations of normal hadronic events.. While it is possible that the events may be seen in a ZDC (Zero Degree Calorimeter) rather than CASTOR, CASTOR is the only LHC detector to provide a general search for Centauro type events.

## 4. PROTON-PROTON EVENTS

CASTOR has observed products of p-p reactions at center of mass energies of 0.9, 2.36, and 7.0 TeV. Figure 2 shows how the increase in the total amount of charge going into the angular range of CASTOR increases as a function of energy. The forward angle cross sections as a function of energy will be useful for improving the cosmic ray shower codes. For elementary particle physics these studies will contribute to such topics as low-x parton dynamics, minimum bias event structure, and diffraction.

As of this writing there has been no evidence of Centauro type events from p-p. Such events would be particularly easy to recognize because of the small number of background products. With Pb-Pb reactions the interesting events must be distinguished from a large background of particles, as predicted by Monte Carlo models such as HIJING.

## 5. STRANGELETS

The quark gluon plasma becomes highly enriched in strange quarks by a process sometimes known as strangeness distillation [5]. For Pb-Pb there are 1248 original u and d quarks. There is sufficient energy to produce a large number of $q\bar{q}$ pairs, but because of the large number of u and d quarks and Pauli blocking, a large fraction of the new $q\bar{q}$ are $s\bar{s}$. The $\bar{s}$ quarks combine with the excess of u and d quarks and escape as kaons. The result of this process is that a considerable fraction of the original u and d quarks are replaced with s quarks.

With a sufficiently small MIT bag constant [5], a part of the original deconfined quark matter that is highly enriched in s quarks remains as essentially a single nucleon containing dozens of quarks. Somehow these objects cool and remain color neutral without the quarks collecting into individual nucleons. Such objects, called strangelets, can be neutral or even negative [5].

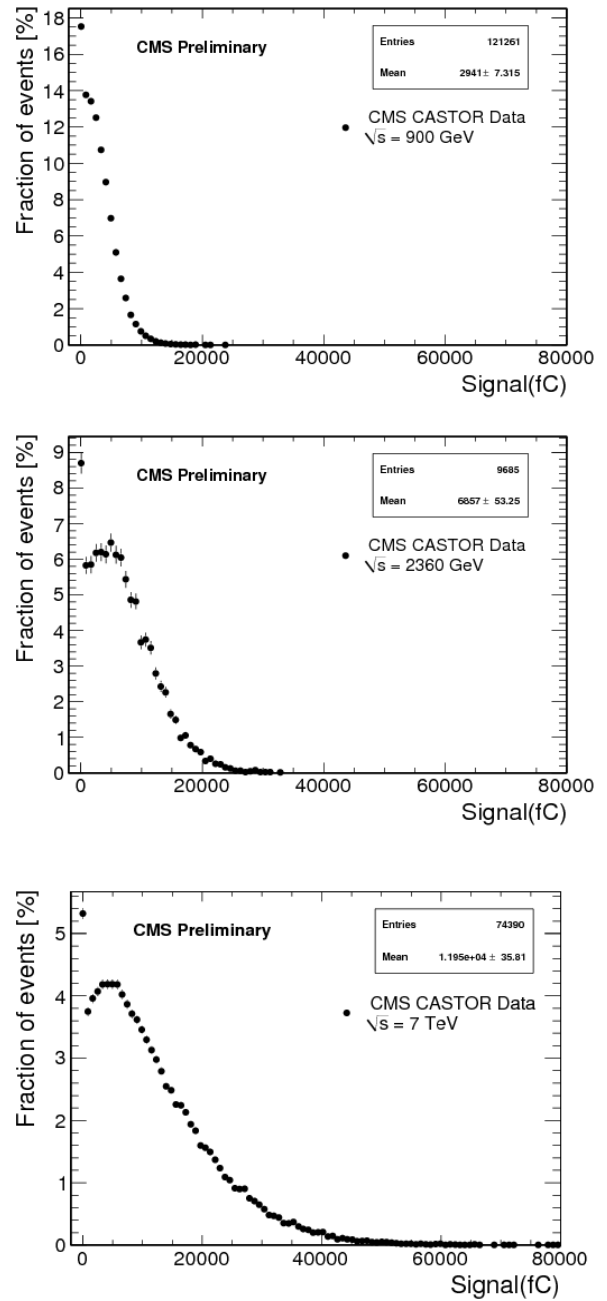

Figure 2: Total charge collected in CASTOR for p-p at $\sqrt{s}$ = 0.9, 2.36, and 7.0 TeV.

Calculations [4] show that 30% of such strangelets would be produced at a large enough angle to fall into the angular range of CASTOR. Increasing the energy above that available at RHIC, $\sqrt{s_{NN}}$ = 200 GeV, reduces the probability of their formation [6]. Strangeness distillation is most effective when the number of $q\bar{q}$ pairs produced is comparable to the number of original u and d quarks.

The RHIC experiment STAR looked at 61 million central Au-Au collisions at $\sqrt{s_{NN}}$ = 200 GeV [7] and set an upper limit of a few times $10^{-6}$ to $10^{-7}$ per collision for strangelets with mass $\gtrsim$ 30 GeV/$c^2$. From this experiment one could conclude that strangelets cannot be made by a quark-gluon plasma and that some other explanation must be found for the cosmic ray events. This conclusion is strengthened by calculations [6] that show that any type of bound system much larger than a single nucleon cannot be produced out of a hot quark gluon plasma. It would be like producing an ice cube in a furnace. The details of the abbreviated discussion above can be found in the references.

We suggest here that the cosmic-ray objects might be hypernuclei with the same quark content as strangelets. Such objects could be produced in high energy Pb-Pb collisions as spectators. For sufficiently non central collisions, some of the spectator matter survives the reaction as actual nuclei. As illustrated in figure 3 for Pb-Pb, the number of spectator neutrons increases as the center to center distance (the impact parameter) between the colliding nuclei increases from zero. Even for completely central collisions there are some spectator neutrons. These are from the outer edge of the Pb nucleus where the neutron density is so low that there is a low probability of interaction. In less central collisions the overlap region becomes the hot quark gluon plasma and the non overlapping part is disintegrated into individual nucleons. Of these spectator nucleons, the neutrons go into the ZDC. With sufficiently large impact parameter the outer parts of the interacting ions are not completely separated into individual nucleons but are left bound into small nuclei. This is seen in figure 3 as a decrease in the number of spectator neutrons for large impact parameter. If partially neutralized spectators are produced by cosmic iron on nitrogen, they would be nuclear spectator fragments infiltrated with strange quarks from the hot overlap region. The spectator nucleus would receive several $K^-$ from the overlap region. If the net charge is sufficiently small the hyper nucleus would go into the ZDC along with the spectator neutrons. The signal from the ZDC would then show points in the region near the red X in figure 3.

Figure 4 shows the beginning of a peripheral collision in the rest frame of the nucleus producing the spectators. In this frame the incoming Pb nucleus at the LHC will have an energy of 4000 TeV in 2010 and $1.5 \times 10^{16}$ eV after the magnets are up to full strength. From the known $K^-$ cross sections it is clear that most of the $K^-$ that enter the spectator will react. The number transferred, and therefore the Z/A of the hypernucleus, depends upon the total number of $K^-$ produced and the direction they take as they leave the interaction region. We have not yet found a $K^-$ angular distribution that would be relevant at these extreme energies.

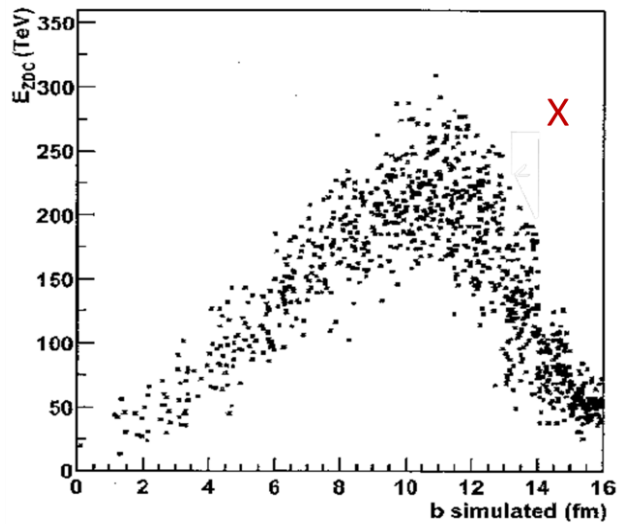

Figure 3: Simulated signals in the ZDC from spectator neutrons with possible hypernuclei, X, as a function of impact parameter.

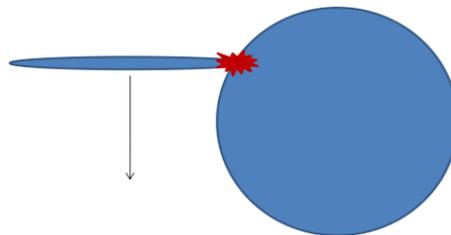

Figure 4: The beginning of a peripheral Pb-Pb collision in the rest frame of one of the nuclei.

Because a spectator fragment is not directly involved in the collision, it leaves the reaction with the same velocity and direction as the nucleus from which it came. It follows that Centauros seen as spectators would not get into CASTOR because they would be at the center of the beam pipe at the CASTOR location. If the spectator Z/A is the same as the beam, 82/208 = 0.4, it will follow the beam through the magnets in the tunnel. If Z/A = 0, it will follow the spectator neutrons, where, for CMS, they are detected in the ZDC, which is located between the incoming and outgoing beam lines at a distance of 140 m from the interaction point.

At CMS a spectator would arrive at a zero degree calorimeter provided if it had received enough negative strange quarks to make Z/A less than about 0.2 and if its lifetime is not too short. The lifetimes of hypernuclei with two lambdas are similar to that of a single lambda, 200-300 ps. If the effective lifetime of a more neutral hypernucleus were also 300 ps, the mean distance to decay would be 26 m at 2.75 TeV/nucleon (LHC) but 10 km at 1000 TeV/nucleon (cosmic rays). Then the objects could easily be seen in cosmic rays but still have problems going the 140 m to the ZDC. (The effective distance is actually only 70 m, because after that there are no more magnets.)

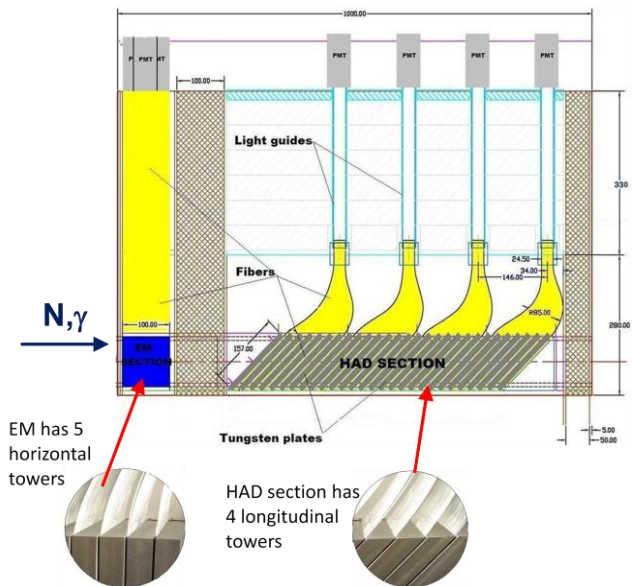

Figure 5: Design of ZDC. Both the EM and the HAD section consist of tungsten plates as the absorbing material. The Čerenkov light from quartz fibers between the metal plates is collected by PMTs at the top.

If a spectator Z/A allows it to get into the ZDC, it would appear as an event with a signal larger than would be expected from the number of spectator neutrons predicted from the impact parameter as measured with larger angle parts of the experiment. In CMS the EM part of the ZDC is divided horizontally into five segments. Figure 5 shows the structure of the CMS ZDC. A spectator with Z/A near 0.2 would have sufficient charge to produce a signal in passing through the EM section, and the EM sector hit would be the one closest to the outgoing beam line.

The total depth of the ZDC is 6.3 nuclear reaction lengths. This is sufficient so that the five point (EM + 4 HAD) distribution of signals from showers of dozens of simultaneous neutrons will have a characteristic shape. A highly penetrating cosmic ray type of particle, whatever its nature, would flatten the curve by leaving considerable energy in the rear HAD towers.

For interactions of cosmic iron with air, much of the iron nucleus becomes spectator matter because iron is larger than nitrogen or oxygen. The cosmic-ray nucleon-nucleon collision energy can be much larger than is available with the LHC, which will result in larger numbers of $K^-$. Nearly neutral spectators may be produced in cosmic rays even if they cannot be produced with beams from the LHC.

The Centauro family includes a wide variety of types of events; it is not likely that a single model would explain them all. While highly strange hypernuclei could possibly be recognized in a zero degree calorimeter, only CASTOR is prepared for a detailed search for the deeply penetrating types of events seen in cosmic rays. The true nature of these events may involve physics concepts that have not yet been proposed by anyone. The first step in the study of such events is their production in a controlled, laboratory situation and the measurement of the cross section for their formation.